\documentclass[12pt]{article} 
\usepackage[sectionbib]{natbib}
\usepackage{array,epsfig,fancyheadings,rotating}
\usepackage[]{hyperref}  
\usepackage{sectsty, secdot}
\usepackage{amssymb,amsfonts}
\usepackage{amsmath}
\DeclareMathOperator{\sign}{sign}
\usepackage{bm}
\usepackage{array}
\usepackage{multirow}
\usepackage{makecell}
\usepackage{color}
\usepackage{geometry}
\usepackage{float}

\sectionfont{\fontsize{12}{14pt plus.8pt minus .6pt}\selectfont}
\renewcommand{\theequation}{\thesection\arabic{equation}}
\subsectionfont{\fontsize{12}{14pt plus.8pt minus .6pt}\selectfont}

\usepackage{amsmath}
\usepackage{amssymb}
\usepackage{amsfonts}
\usepackage{multirow}
\usepackage{amsthm}

\newtheorem{theorem}{Theorem}
\newtheorem{lemma}{Lemma}
\newtheorem{corollary}{Corollary}
\newtheorem{proposition}{Proposition}
\theoremstyle{definition}
\newtheorem{definition}{Definition}
\newtheorem{example}{Example}

\newtheorem{condition}{Condition}
\usepackage{xr}
\newcommand{\corr}{{\rm Corr}}
\newcommand{\cov}{{\rm Cov}}
\newcommand{\var}{{\rm Var}}

\begin{document}


\renewcommand{\baselinestretch}{2}

\markright{ \hbox{\footnotesize\rm Statistica Sinica
}\hfill\\[-13pt]
\hbox{\footnotesize\rm
}\hfill }

\markboth{\hfill{\footnotesize\rm FEI XUE AND ANNIE QU} \hfill}
{\hfill {\footnotesize\rm SEMI-STANDARD PARTIAL COVARIANCE SELECTION} \hfill}

\renewcommand{\thefootnote}{}
$\ $\par


\fontsize{12}{14pt plus.8pt minus .6pt}\selectfont \vspace{0.8pc}
\centerline{\large\bf SEMI-STANDARD PARTIAL COVARIANCE}
\vspace{2pt} \centerline{\large\bf VARIABLE SELECTION WHEN}
\vspace{2pt} \centerline{\large\bf IRREPRESENTABLE CONDITIONS FAIL}
\vspace{.4cm} \centerline{Fei Xue and Annie Qu} \vspace{.4cm} \centerline{\it
Purdue University and University of California Irvine} \vspace{.55cm} \fontsize{9}{11.5pt plus.8pt minus
.6pt}\selectfont


\begin{quotation}
\noindent {\it Abstract:}
{Traditional variable selection methods could fail to be sign consistent when irrepresentable conditions are violated.
	This is especially critical in high-dimensional settings when the number of predictors exceeds the sample size. In this paper, we propose a new semi-standard partial covariance (SPAC) approach which is capable of reducing correlation effects from other covariates while fully capturing the magnitude of coefficients. The proposed SPAC is effective in choosing covariates which have direct effects on the response variable, while eliminating the predictors which are not directly associated with the response but are highly correlated with the relevant predictors. We show that the proposed SPAC method with the Lasso penalty or the smoothly clipped absolute deviation (SCAD) penalty possesses strong sign consistency in high-dimensional settings. Numerical studies and a post-traumatic stress disorder data application also confirm that the proposed method outperforms the existing Lasso, adaptive Lasso, SCAD, Peter--Clark-simple algorithm, and factor-adjusted regularized model selection methods when the irrepresentable conditions fail.}\\
\noindent {\it Key words and phrases:}
  Irrepresentable condition, Lasso, model selection consistency, partial correlation, smoothly clipped absolute deviation.
\par
\end{quotation}\par

\def\thefigure{\arabic{figure}}
\def\thetable{\arabic{table}}

\renewcommand{\theequation}{\thesection.\arabic{equation}}

\fontsize{12}{14pt plus.8pt minus .6pt}\selectfont

\setcounter{equation}{0} 

\section{Introduction}

Variable selection is an important model-building tool for selecting covariates relevant to the response variable, which is fundamental for the construction of a sparse model when the number of relevant covariates is much smaller than the total number of observed covariates. This is especially crucial under high-dimensionality where the number of covariates far exceeds the number of observations. For high-dimensional data, traditional regularization variable selection methods (\citealp{Ti}; \citealp{FL}; \citealp{zou2005regularization}; \citealp{yuan2006model}; \citealp{Zou}; \citealp{candes2007dantzig}; \citealp{Zh}) are effective in achieving model selection and parameter estimation simultaneously 
under irrepresentable conditions (\citealp{ZY}; \citealp{fan2011nonconcave}; \citealp{kim2008smoothly}), which assume that correlations between relevant and irrelevant covariates are relatively weak compared with correlations among relevant covariates.

\lhead[\footnotesize\thepage\fancyplain{}\leftmark]{}\rhead[]{\fancyplain{}\rightmark\footnotesize\thepage}

However, the irrepresentable conditions could fail whether the dimension is high or not.
For example, in mediation analysis for identifying mediators which transmit effects from an exposure factor to an outcome variable, spurious mediators (irrelevant covariates) could be strongly correlated with the exposure factor and true mediators (relevant covariates) (\citealp{jerolon2018causal}; \citealp{chen2017high}; \citealp{imai2013identification}). Although some modified model selection methods have been proposed to incorporate strongly correlated covariates, they either do not possess variable selection consistency (\citealp{wang2014combination}; \citealp{maier2017fast}; \citealp{hilafu2017sufficient}; \citealp{buhlmann2013correlated}), or impose a more restrictive condition such as knowing the true number of relevant covariates \citep{javanmard2013model}. In particular, several existing methods (\citealp{sharma2013consistent}; \citealp{fu2014group}; \citealp{zeng2012group}; \citealp{huang2016mnet}) tend to group and select highly correlated relevant and irrelevant predictors together. \citet{jia2015preconditioning} propose to transform the design matrix so that the irrepresentable conditions are satisfied. However, the error terms are no longer independent from each other after transformation. More importantly, model-based transformation loses its original interpretation in practice.

Under high-dimensional settings \citep{fan2018discoveries}, sure independence screening \citep{fan2008sure} screens out variables through marginal correlations between the response and covariates. However, the marginal correlations
between irrelevant covariates and the response could increase when the irrelevant covariates are strongly correlated with relevant covariates, which may reduce the effectiveness of the sure independence screening.
The Peter--Clark-simple (PC-simple) algorithm \citep{buhlmann2010variable} was developed to screen variables via partial correlation to solve the correlation problem.
Moreover, \citet{cho2012high} generalize the partial correlation to the tilted correlation, and \cite{li2016} and \cite{jin2014optimality} incorporate inter-feature correlations to improve detection of marginally weakly associated covariates. In addition, \citet{bradic2016randomized} proposes a  subsample bootstrap aggregation approach to circumvent the irrepresentable conditions, and \citet{fan2018factor} develop the factor-adjusted regularized model selection (Farm-Select) method to  
decorrelate highly-correlated covariates. 

The partial correlation approach measures each individual covariate effect after removing other covariate effects (\citealp{PWZZ}; \citealp{buhlmann2010variable}; \citealp{li2015variable}; \citealp{tang2017testing}). However, the range of partial correlation is bounded between minus one and one, and therefore the partial correlation may not fully capture strong signals of some relevant covariates.
This motivates us to develop a new semi-standard partial covariance (SPAC) approach to fully use the magnitude of signal strength. 
The proposed SPAC is more powerful than partial correlation in identifying relevant covariates.

Compared with traditional regularization methods, the proposed method encourages selecting covariates which have direct effects on the response variable, while
discouraging the selection of irrelevant covariates which are strongly correlated with relevant covariates.
We demonstrate estimation consistency and variable selection consistency for the proposed SPAC method with the Lasso penalty (SPAC-Lasso) or the SCAD penalty (SPAC-SCAD). 
The proposed method can handle both fixed-dimensional settings and high-dimensional settings 
when relevant and irrelevant covariates are highly correlated with each other. 



Our work has the following contributions. First, the proposed variable selection approach can mitigate the bias of model selection caused by the violation of irrepresentable conditions for the Lasso or the SCAD method.
We show that the proposed SPAC-Lasso and SPAC-SCAD are still sign consistent and especially effective when correlations between relevant and irrelevant covariates are higher than correlations among relevant covariates. Second, the proposed SPAC is more effective in acquiring signal strength and thus more powerful in selecting relevant predictors than traditional partial correlation. Numerical studies confirm that the proposed method outperforms traditional penalty-based variable selection methods, the PC-simple algorithm and the Farm-Select method for highly dependent covariates.

This paper is organized as follows. Section 2 provides the model framework for the variable selection problem. Section 3 introduces the SPAC and presents the proposed methodology. Section 4 establishes theoretical properties of the SPAC-Lasso and SPAC-SCAD.  Section 5 discusses the implementation of the proposed method. Sections 6 presents various simulation studies. Section 7 illustrates a real data application to post-traumatic stress disorder (PTSD) of African Americans.

%


\setcounter{equation}{0} 
\section{Model framework and notation}\label{notation}

We formulate the variable selection problem under a linear regression setting,
\begin{equation}\label{setting}
\bm{y}=\bm{X}\bm{\beta}+\bm{\varepsilon},
\end{equation}
where $\bm{y}=(y_1, \dots, y_n)^T$ consists of samples for the response variable $Y$, 
$\bm{X}=(x_{ij})$ is a $n\times p$ random design matrix, $\bm{\beta}=(\beta_1, \dots, \beta_{p})^T$ is a coefficient vector, and the noise vector $\bm{\varepsilon} \sim N_n(\bm{0}, \sigma_{\varepsilon}^2\bm{I}_n)$ is uncorrelated with $\bm{X}$. Let  $\bm{x}_j$ 
be the $j\text{th}$ column ($j\text{th}$ covariate) 
of $\bm{X}$
for each $j=1, \ldots, p$.
Without loss of generality, we assume that each column is standardized from independently and identically distributed samples, that is, $\bm{x}_j^T\bm{x}_j=n$ and mean $\sum_{i=1}^{n} x_{ij}=0$ for $j=1, \dots. n$.
Then each row of $\bm{X}$
is identically distributed from a $p$-dimensional random vector $\bm{\mathcal{X}}=(X_1, \dots, X_{p})^T$
with mean $\bm{0}$ and positive-definite covariance matrix $\bm{C}_{p\times p}$ whose diagonal elements are all $1$'s. In addition, we assume that the response variable is standardized with $\sum_{i=1}^{n}y_i=0$, and thus the intercept can be omitted.

Here we assume that the linear model in (\ref{setting}) is sparse, where most covariates have zero coefficients and are irrelevant to the response $Y$.
That is, only the first $q$ covariates in $\bm{X}$ have non-zero coefficients and are relevant to the response variable, and let $\beta_i=0$ if and only if $i>q$. In addition, we let $\bm{\Sigma}=\cov(Y, X_1, \dots, X_{p})$, and $\bm{\Sigma}^{-1}=(\sigma^{ij})$, where $i,j \in \{Y,1,2,\dots,p\}.$


Under the sparsity assumption,
the penalized least squares regression methods (\citealp{Ti}; \citealp{Fu}) select variables
through minimizing the penalized least squares function
\vspace{-1mm}
\begin{equation}\label{L0}
L(\bm{\beta})=\frac{1}{2}\|\bm{y}-\bm{X}\bm{\beta}\|^2+\sum_{j=1}^{p}p_{\lambda} (\beta_j),
\end{equation}
where $\|\cdot\|$ represents the Euclidean norm, $p_{\lambda}(\cdot)$ is a penalty function, and $\lambda$ is a tuning parameter. Here the $p_{\lambda}(\beta_j)$ could be the Lasso, adaptive Lasso, or SCAD penalty, which have the forms $p_{Lasso, \lambda}(\beta_j)=\lambda |\beta_j|$, $p_{ALasso, \lambda}(\beta_j)=\lambda|\beta_j|/|\hat{\beta}_{0j}|$, and
\begin{equation}\label{SCAD}
p_{SCAD, \lambda}(\beta_j)=\left\{\begin{array}{ll}
\lambda |\beta_j|&\text{if } 0\le |\beta_j|\le\lambda\\
\frac{a\lambda |\beta_j|-0.5(|\beta_j|^2+\lambda^2)}{a-1}&\text{if } \lambda<|\beta_j|\le a \lambda\\
\frac{\lambda^2(a^2-1)}{2(a-1)}&\text{if } |\beta_j|>a \lambda,
\end{array}\right.
\end{equation}
where ``ALasso'' represents the adaptive Lasso penalty, 
$a>2$, and $\hat{\beta}_{0j}$ is an initial estimator of $\beta_j$. 



\setcounter{equation}{0} 
\section{A New Variable Selection Method}

In this section, we propose a semi-standard partial covariance (SPAC) variable selection approach to achieve selection consistency when the original irrepresentable conditions (\citealp{ZY}; \citealp{fan2011nonconcave}) fail, that is, there exist 
strong correlations between relevant and irrelevant covariates. The proposed SPAC is able to capture the relationship between a relevant covariate and the response variable conditional on other covariate effects, as we derive this SPAC 
from the notion of partial correlation.
For each $j=1, \dots, p$, let $\rho_j=\corr (\varepsilon_Y, \varepsilon_j)$ 
be the partial correlation 
between the response $Y$ and covariate $X_j$, 
where $\varepsilon_Y$ and $\varepsilon_j$ are the residuals of linear regression models with $Y$ and $X_j$ as responses, respectively, and $X_{-j}=\{X_k: k=1, \dots, j-1, j+1, \dots, p\}$ as predictors.

Under the normality assumption
\begin{equation}\label{Normal}
(Y, X_1, \dots, X_{p})^T \sim N_{p+1}(\bm{0}, \bm{\Sigma}),
\end{equation}
it is well-known that $\rho_j=\corr(Y,X_j\mid X_{-j})$ 
\citep{baba2004partial}, indicating that the partial correlation 
measures the linear relationship between $Y$ and $X_j$ conditional on other covariates. Moreover, non-zero partial correlations correspond to relevant covariates,
while zero partial correlations correspond to irrelevant ones. 

However, partial correlation is unable to fully capture signal strength, which is the magnitude of $\beta_j$, due to its bounded range. To overcome this limitation, we propose the following SPAC and provide the association between the SPAC and partial correlation in Lemma \ref{lemma1}.
\begin{definition}\label{gamma}
	The semi-standard partial covariance (SPAC) between the response $Y$ and covariate $X_j$ is $\gamma_j=\beta_j/ d_{jj}^{1/2}$
	for $j=1,\dots,p$, where 
	$d_{jj}$ is the $j$th diagonal element of the precision matrix $\bm{D}=\bm{C}^{-1}$.
\end{definition}
The exponent $1/2$ of $d_{jj}$ in Definition \ref{gamma} ensures that $\gamma_j$ does not depend on the scale of $X_j\mid X_{-j}$ by the following lemma.
\begin{lemma}\label{lemma1}
	Let $s_j = \{\var(Y\mid X_{-j})\}^{1/2}$ for each $j=1, \dots, p$.
	Under the normality assumption (\ref{Normal}), we have 
	$$
	\gamma_j = \rho_j s_j =\frac{\cov (Y, X_j \mid X_{-j})}{\{\var (X_{j} \mid X_{-j})\}^{1/2}},\ \ \
	s_j^2=\frac{1/\sigma^{YY}}{1-\rho_j^2}=\frac{\beta_j^2}{d_{jj}}+\sigma_\varepsilon^2.$$
\end{lemma}
By definition, $\gamma_j=0$ if and only if $\beta_j=0$ for each $j=1, \dots, p$, implying that we can select relevant covariates through identifying non-zero SPACs.
Lemma \ref{lemma1} shows that the SPAC is equivalent to the multiplication of the partial correlation and $s_j$ under the normality assumption.
Moreover, the proposed $\gamma_j$ standardizes the partial covariance $\cov (Y, X_j \mid X_{-j})$ by $\{\var (X_{j} \mid X_{-j})\}^{1/2}$, instead of both $s_j$ and $\{\var (X_{j} \mid X_{-j})\}^{1/2}$ in the partial correlation, which is the reason that
we refer to $\gamma_j$ as ``semi-standard'' partial covariance.
We involve $s_j$ in the SPAC, since
$s_j$ is an increasing function of partial correlation $\rho_j$ and also incorporates the magnitude of coefficient $\beta_j$ as indicated in Lemma \ref{lemma1}.
Therefore, the proposed SPAC
is able to fully capture the signal strength of relevant predictors while removing effects from other covariates. 


We illustrate SPAC and compare it with partial correlation from a geometric perspective using a toy example. 
Let $\bm{y}=\beta_1 \bm{x}_1+\beta_2 \bm{x}_2+\bm{\varepsilon}$ with $\beta_1\ne 0$ and $\beta_2= 0$, that is, $\bm{x}_1$ is relevant but $\bm{x}_2$ is irrelevant. We also assume that $\bm{x}_1$ and $\bm{x}_2$ are correlated. By definition, $\gamma_1\ne 0$ and $\gamma_2=0$. 

\begin{figure}
	\begin{center}
		\resizebox{160.632pt}{207pt}{\includegraphics{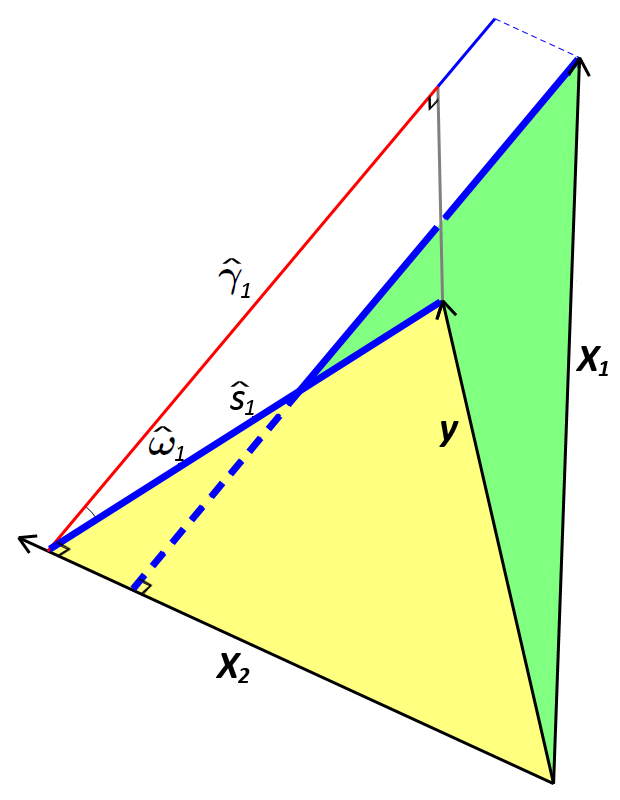}} 
		\hspace{15mm}\resizebox{160.632pt}{207pt}{\includegraphics{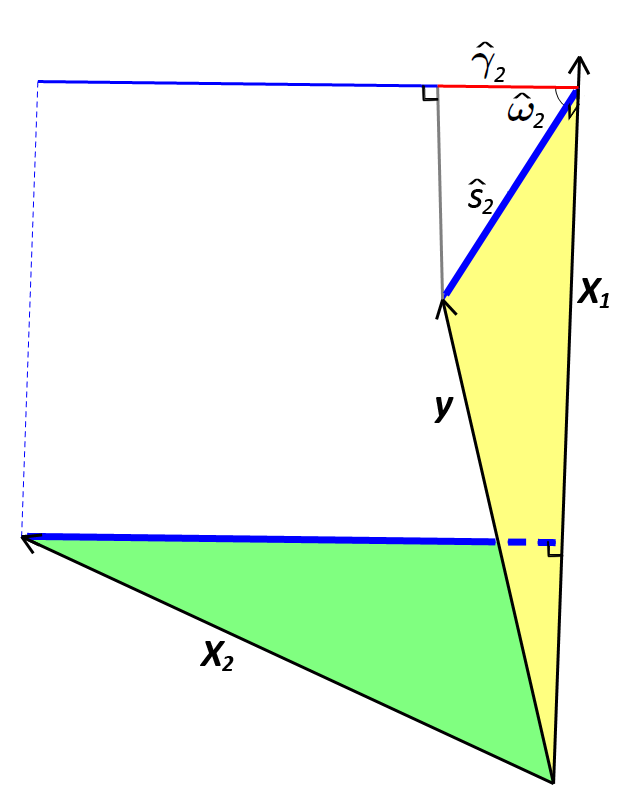}}
		\caption{Illustrations of SPAC and partial correlation when $X_1$ and $X_2$ are correlated.}\label{Example}
	\end{center}
\end{figure}

We plot the relationships of $\bm{x}_1$, $\bm{x}_2$, and $\bm{y}$ in Figure \ref{Example}. As shown in the left graph, $\hat{\omega}_1$ is the angle between the two bold blue lines which represent residuals of projections from $\bm{y}$ and $\bm{x}_1$ onto $\bm{x}_2$. Then $\hat{\rho}_1=\cos (\hat{\omega}_1)$ is the sample partial correlation based on samples in $\bm{x}_1$, $\bm{x}_2$, and $\bm{y}$. The length of the bold blue line for residuals of $\bm{y}$ is a sample estimator of $s_1$, denoted by $\hat{s}_1$. By Lemma \ref{lemma1}, $\hat{\gamma}_1=\hat{s}_1\cos (\hat{\omega}_1)$ is a sample estimator for $\gamma_1$, which is also the projection from residuals of $\bm{y}$ onto residuals of $\bm{x}_1$, represented by the red line in the left graph. 

Similarly in the right graph of Figure \ref{Example}, $\hat{\rho}_2=\cos (\hat{\omega}_2)$ and $\hat{\gamma}_2$ are the sample partial correlation and sample SPAC for $\bm{x}_2$, respectively. The $\hat{\gamma}_2$ is not exactly zero due to sample variation. The differences between sample SPACs and sample partial correlations come from $\hat{s}_1$ and $\hat{s}_2$. As shown in Figure \ref{Example},  $\hat{s}_2$ is just the sample variance of the error term, while $\hat{s}_1$ contains the error variation and increases as signal coefficient $\beta_1$ increases, implying that $\hat{s}_1$ should be larger than $\hat{s}_2$. Therefore, SPAC is more effective in distinguishing relevant covariates from irrelevant covariates than partial correlation.


Compared with coefficients $\bm{\beta}$, the SPAC takes account of correlation effects from other covariates.
Specifically, since $1/d_{jj}^{1/2}=\{\var(X_j  \mid  X_{-j})\}^{1/2}=(1-R_j^2)^{1/2}$  (\citealp{La}; \citealp{raveh1985use}), the SPAC for covariate $X_j$ is
\begin{equation*}
\gamma_j = \beta_j \left\{\var(X_j  \mid  X_{-j})\right\}^{1/2}=\beta_j\left(1-R_j^2\right)^{1/2},
\end{equation*}
where $R_j$ is the coefficient of multiple correlation between $X_j$ and all other covariates. When $X_j$ is independent of other covariates, $\gamma_j$ is the same as $\beta_j$.  
On the other hand, when $X_j$ is correlated with other covariates, the SPAC mitigates correlation effects from other covariates via multiplying $\beta_j$ by $(1-R_j^2)^{1/2}$. 
Thus, we propose to focus on estimating SPAC $\gamma_j$ instead of coefficient $\beta_j$ to achieve model selection consistency for data with strong correlations between irrelevant covariates and relevant covariates.



Specifically, we replace coefficient $\beta_j$ in the penalized least squares function (\ref{L0}) with $\hat{d}_{jj}^{1/2}\gamma_j$ for each $j=1, \dots, p$, and estimate $\bm{\gamma}=(\gamma_1, \dots, \gamma_{p})^T$ by minimizing
\begin{eqnarray} \label{generalLoss}
L(\bm{\gamma},\hat{\bm{d}})&=&\frac{1}{2}\|\bm{y}-\sum_{j=1}^{p} \bm{x}_j\hat{d}_{jj}^{1/2}\gamma_j\|^2+\sum_{j=1}^{p}p_{\lambda}(\gamma_j)\hat{d}_{jj},
\end{eqnarray}
where 
$\hat{\bm{d}}=(\hat{d}_{11},\dots,\hat{d}_{pp})^T$ is a consistent estimator of diagonal elements $\bm{d}=(d_{11},\dots,d_{pp})^T$. 
Substituting $\bm{\beta}$ by $\bm{\gamma}$, we obtain a new matrix $\bm{X}^*=(\bm{x}_1\hat{d}_{11}^{1/2}, \dots,\bm{x}_{p}\hat{d}_{p p}^{1/2})$ which serves similarly as a design matrix for $\bm{\gamma}$.  The squared Euclidean norm of the $j\text{th}$ column in $\bm{X}^*$ is $\hat{d}_{jj} \bm{x}_j^T \bm{x}_j=\hat{d}_{jj} n$ for $j=1, \dots, p$, which leads to different weights on penalization 
for different covariates. 
However, the SPAC of each covariate could be equally important. To avoid unequal weighting, we reweight the penalization term via multiplying the penalty $p_{\lambda}(\gamma_j)$ in (\ref{generalLoss}) by $\hat{d}_{jj}$ for each $j=1, \dots, p$. Consequently, the proposed SPAC estimator is
\begin{equation*}
\hat{\bm{\gamma}}= \underset{\bm{\gamma}}{\text{argmin}}\ L(\bm{\gamma},\hat{\bm{d}}),
\end{equation*}
and the corresponding estimator for the coefficients is $\hat{\bm{\beta}}=(\hat{d}_{11}^{1/2}\hat{\gamma}_1,  \dots,  \hat{d}_{p p}^{1/2}\hat{\gamma}_{p})^T$. 

We adopt the Lasso, adaptive Lasso, and SCAD penalty functions, respectively, to shrink SPACs in the following Examples \ref{ELasso}--\ref{ESCAD}, and refer to the corresponding estimators as SPAC-Lasso, SPAC-ALasso, and SPAC-SCAD, respectively. 
We compare these estimators with the original Lasso, adaptive Lasso, and SCAD estimators in Sections \ref{simulation} and \ref{real}.


\begin{example} \label{ELasso}
	If we use Lasso penalty, the penalized loss function in (\ref{generalLoss}) becomes
	\begin{eqnarray}\label{lossLasso}
	L_{Lasso}(\bm{\gamma},\hat{\bm{d}})&=&\frac{1}{2}\| \bm{y}-\sum_{j=1}^{p}  \bm{x}_j \hat{d}_{jj}^{1/2}\gamma_j \|^2+\lambda\sum_{j=1}^{p}\hat{d}_{jj}|\gamma_j|.
	\end{eqnarray}
	Accordingly, the proposed estimator with Lasso penalty (SPAC-Lasso) is 
	\begin{equation*}
	\hat{ \bm{\gamma}}_{Lasso} = \underset{ \bm{\gamma}}{\text{\rm argmin}}\ L_{Lasso}( \bm{\gamma},\hat{ \bm{d}}).
	\end{equation*}
\end{example}
\begin{example} \label{EALasso}
	Suppose that $\hat{ \bm{\gamma}}_{0}=(\hat{\gamma}_{01}, \dots, \hat{\gamma}_{0p})^T$ is a consistent initial estimator for $ \bm{\gamma}$. The objective function for the SPAC method with the adaptive Lasso penalty (SPAC-ALasso) is
	\begin{equation} \label{aLassoloss}
	L_{ALasso}( \bm{\gamma},\hat{ \bm{d}})=\frac{1}{2}\|\bm{y}-\sum_{j=1}^{p}  \bm{x}_j\hat{d}_{jj}^{1/2}\gamma_j\|^2+\lambda\sum_{j=1}^{p} \hat{d}_{jj}\frac{|\gamma_j|}{|\hat{\gamma}_{0j}|^{\mu}},
	\end{equation}
	where $\mu >0$ is a tuning parameter. The corresponding SPAC-ALasso estimator is 
	\begin{equation*}
	\hat{\bm{\gamma}}_{ALasso} = \underset{\bm{\gamma}}{\text{\rm argmin}}\ L_{ALasso}(\bm{\gamma},\hat{\bm{d}}).
	\end{equation*}
\end{example}
\begin{example} \label{ESCAD}
	Similarly, the objective function for the proposed SPAC method with the SCAD penalty (SPAC-SCAD) is 
	\begin{equation}\label{lossSCAD}
	L_{SCAD}(\bm{\gamma},\hat{\bm{d}})=\frac{1}{2}\|\bm{y}-\sum_{j=1}^{p}  \bm{x}_j\hat{d}_{jj}^{1/2}\gamma_j\|^2+n\sum_{j=1}^{p}p_{SCAD,\lambda}(\gamma_j)\hat{d}_{jj},
	\end{equation}
	where $p_{SCAD, \lambda}(\cdot)$ is defined in (\ref{SCAD}), and the corresponding SPAC-SCAD estimator is 
	\begin{equation*}
	\hat{\bm{\gamma}}_{SCAD} = \underset{\bm{\gamma}}{\text{\rm argmin}}\ L_{SCAD}(\bm{\gamma},\hat{\bm{d}}).
	\end{equation*}
\end{example}


\setcounter{equation}{0}
\section{Consistency Theory}\label{theory}
In this section, we demonstrate the asymptotic properties of the proposed SPAC-Lasso and SPAC-SCAD estimators, and provide examples satisfying  conditions for consistency of the proposed method. Although Lemma \ref{lemma1} is under the normality assumption, we do not require this normality assumption in the following  subsection.

\subsection{Consistency under high dimensionality}

In this subsection, we establish the variable selection consistency and estimation consistency of the SPAC-Lasso and SPAC-SCAD under high dimensionality, where $p=p_n$, $q=q_n$, and $\bm{C}=\bm{C}_n$ increase as $n$ grows.
Similar results for fixed dimensions of $p$ and $q$ are provided in Section S2 
in the supplementary material.
For the high-dimensional settings, the Lasso, adaptive Lasso, and SCAD methods require correlations between relevant and irrelevant covariates being relatively small compared with correlations among relevant covariates to achieve variable selection consistency (\citealp{ZY}; \citealp{huang2008adaptive}; \citealp{kim2008smoothly}; \citealp{fan2011nonconcave}). The proposed SPAC approach mitigates correlation effects from other covariates to achieve model selection consistency when relevant and irrelevant covariates are strongly correlated and the original irrepresentable conditions fail. 

Following similar notation as in \citet{ZY}, let $\hat{\bm{\gamma}} =_s \bm{\gamma}$ if and only if $\sign (\hat{\bm{\gamma}}) = \sign(\bm{\gamma})$, 
and an estimator $\hat{\bm{\gamma}}$ is \textbf{strongly sign consistent} if there exists a tuning parameter $\lambda_n$, a function of $n$, such that $$\lim\limits_{n\rightarrow\infty}P\left\{\hat{\bm{\gamma}}(\lambda_n)=_s\bm{\gamma}\right\}=1,$$ 
where the $\lambda_n$ is independent of the data.

To show the sign consistency of the proposed method, we define the following notations. Let $\bm{X} (1)$ and $\bm{X} (2)$ be the first $q_n$ and remaining $p_n-q_n$ columns in $\bm{X}$, respectively, such that $\bm{X} (1)$ contains relevant covariates, and $\bm{X}(2)$ consists of irrelevant covariates. Let $\hat{\bm{C}}_n=\bm{X}^T\bm{X}/n$ be the sample covariance matrix of $\bm{X}$, with diagonal elements all $1$'s since the covariates are standardized as mentioned in Section \ref{notation}. Thus, $\hat{\bm{C}}_n$ and the true covariance matrix $\bm{C}_n$ are both correlation matrices, and can be partitioned into blocks 
$$\hat{\bm{C}}_n=\begin{pmatrix}
\hat{\bm{C}}^{11}_n&\hat{\bm{C}}^{12}_n\\\hat{\bm{C}}^{21}_n& \hat{\bm{C}}^{22}_n
\end{pmatrix}, \ \ \ 
\bm{C}_n=\begin{pmatrix}
\bm{C}^{11}_n&\bm{C}^{12}_n\\\bm{C}^{21}_n& \bm{C}^{22}_n
\end{pmatrix},$$
according to $\bm{X} =(\bm{X}(1), \bm{X}(2))$.
Similarly, we partition $\bm{\gamma}$ into $\bm{\gamma}(1) = (\gamma_1, \dots, \gamma_{q})^T$ and $\bm{\gamma}(2) = (\gamma_{q+1}, \dots, \gamma_{p})^T$, representing relevant and irrelevant coefficients of SPACs, respectively.
In addition, we define conditions for the proposed SPAC-Lasso and SPAC-SCAD as follows.

\begin{condition}[\textbf{Irrepresentable condition for SPAC-Lasso}]\label{irepLasso}
	There exists a positive constant $\eta$ such that
	\begin{equation*}\label{weightcon}
	\left\|\bm{V}(2)\hat{\bm{C}}^{21}_n(\hat{\bm{C}}^{11}_n)^{-1}\bm{V}(1)^{-1}\sign\{\bm{\beta}(1)\}\right\|_{\infty}\le 1-\eta, 
	\end{equation*}
	where $\|\cdot\|_{\infty}$ represents the infinity norm of a matrix,
	and $\bm{V}(1)$ and $\bm{V}(2)$ are diagonal matrices $\text{diag}\{1/d_{11}^{1/2},\ldots,1/d_{qq}^{1/2}\}$ and $\text{diag}\{1/d_{q+1 q+1}^{1/2},\ldots,1/d_{pp}^{1/2}\}$, respectively. 
\end{condition}


\begin{condition}[\textbf{Irrepresentable condition for SPAC-SCAD}]\label{irepSCAD}
	There exists a positive constant $\eta$ such that
	\begin{equation*}
	\mathcal{P}_{\lambda^*_n}\left(h_{\min}\right)\left\|\bm{V}(2)\hat{\bm{C}}^{21}_n(\hat{\bm{C}}^{11}_n)^{-1}\bm{V}(1)^{-1}\right\|_{\infty}\le 1-\eta,
	\end{equation*}
	where $\mathcal{P}_{\lambda^*_n}(\cdot)=p'_{SCAD,\lambda^*_n}(\cdot)/\lambda^*_n$, $h_{\min}=\min_{1\le j \le q_n} |\beta_j|/2$, and $\lambda^*_n=\lambda_n \max_{1\le j \le q_n}d_{jj}^{1/2}$.
	
\end{condition}

Condition \ref{irepLasso} is required for the sign consistency of the SPAC-Lasso , while Condition \ref{irepSCAD} is required for the SPAC-SCAD under high-dimensional settings. Condition \ref{irepSCAD} is weaker than Condition \ref{irepLasso} when signals are strong since the SCAD penalty gradually levels off. The above two conditions are modified from the original irrepresentable conditions proposed in \citet{ZY} and \citet{fan2011nonconcave} for Lasso and SCAD, respectively. However, 
the proposed 
Conditions \ref{irepLasso} and \ref{irepSCAD} could still hold for cases where the original irrepresentable conditions fail. We illustrate this with examples in Section \ref{examples}.

\begin{condition}\label{dime}
	For some positive constants $0<\kappa_0, \kappa_2<1/2$ and $\kappa_1>0$, $\log p_n=O(n^{1-2\kappa_0})$, $q_n = O(n^{\kappa_2})$, $h_{\min}\ge (\log p_n/n)^{1/2}$, and $p_n\ge n^{\kappa_1}$.
\end{condition}

Condition \ref{dime} allows the number of covariates to grow exponentially, but requires a lower bound of signal strength similarly as in \citet{fan2013asymptotic} and \citet{zheng2014high}). The requirement $p_n\ge n^{\kappa_1}$ comes from \citet{cai2011constrained} to ensure the consistency of the constrained $L_1$-minimization estimator (CLIME) \citep{cai2011constrained} which is adopted in the following theorems. We let $\hat{\bm{d}}$ be the diagonal elements of the CLIME. Then, $\hat{\bm{d}}$ is consistent under some regularity conditions and a sparsity assumption of the precision matrix 
\citep{cai2011constrained}.

Following notations in \cite{cai2011constrained}, we model the sparsity of the precision matrix $\bm{D}$ through defining
\begin{eqnarray}\label{sparse}
\mathcal{G}_u(K_{p_n}, M_{p_n})=\left\{\bm{D}: \underset{1\le j\le p_n}{\max} \sum_{i=1}^{p_n}|d_{ij}|^u\le K_{p_n}, \|\bm{D}\|_1\le M_{p_n}\right\},\ \ \ \ \ \ \
\end{eqnarray}
where $0\le u <1$, $K_{p_n}$ and $M_{p_n} $ are positive and allowed to increase as $n$ grows. 
We consider data with precision matrices $\bm{D}\in \mathcal{G}_u(K_{p_n}, M_{p_n})$ throughout this subsection. Details of other regularity Conditions 4, 5, and 6 
are provided in Section S1 
in the supplementary material. Also, proofs for the following theorems are provided in Section S4 
in the supplementary material.
\begin{theorem}
	\label{Lassolargesuf}
	Let $\hat{\bm{d}}$ be diagonal elements of the CLIME of $\bm{D}$. 
	If Conditions \ref{dime}, 4, 
	and 5 
	are satisfied, 
	and Condition \ref{irepLasso} holds with probability at least $1-O(n^{-\delta})$, 
	then we have the following properties for the minimization of $L_{Lasso}(\bm{\gamma}, \hat{\bm{d}})$ in (\ref{lossLasso}) with probability at least $1-O(n^{-\delta})$.
	
	(1) Strong sign consistency: There exists a strict local minimizer $\hat{\bm{\gamma}}_{Lasso}$ such that 
	$\hat{\bm{\gamma}}_{Lasso}=_s\bm{\gamma}$.
	
	(2) Estimation consistency: The corresponding estimator of coefficients $\hat{\bm{\beta}}=\hat{\bm{V}}^{-1}\hat{\bm{\gamma}}_{Lasso}$ satisfies
	$$\|\hat{\bm{\beta}}-\bm{\beta}\|_{\infty}= O\left\{(\log p_n/n)^{1/2}\right\}.$$
\end{theorem}

\begin{theorem}
	\label{SCADlargesuf}
	Let $\hat{\bm{d}}$ be diagonal elements of the CLIME of $\bm{D}$. If Conditions \ref{dime}, 5, and 6
	are satisfied, and Condition \ref{irepSCAD} holds with probability at least $1-O(n^{-\delta})$, then we have the following properties for the minimization of $L_{SCAD}(\bm{\gamma}, \hat{\bm{d}})$ in (\ref{lossSCAD}) with probability at least $1-O(n^{-\delta})$.
	
	(1) Strong sign consistency: There is a strict local minimizer $\hat{\bm{\gamma}}_{SCAD}$ such that 
	$\hat{\bm{\gamma}}_{SCAD}=_s\bm{\gamma}$.
	
	(2) Estimation consistency: The corresponding estimator of coefficients $\hat{\bm{\beta}}=\hat{\bm{V}}^{-1}\hat{\bm{\gamma}}_{SCAD}$ satisfies
	$$\|\hat{\bm{\beta}}-\bm{\beta}\|_{\infty}= O\left\{(\log p_n/n)^{1/2}\right\}.$$
\end{theorem}

Theorems \ref{Lassolargesuf} and \ref{SCADlargesuf} state that, even though the number of covariates increases exponentially, the proposed SPAC-Lasso and SPAC-SCAD are able to select the true model
with probability tending to $1$ under Conditions \ref{irepLasso} and \ref{irepSCAD}, respectively. Moreover, the estimators for coefficients based on  
SPAC-Lasso and SPAC-SCAD both
converge to the true $\bm{\beta}$. 

Note that Condition \ref{irepLasso} could still be valid even when the original  irrepresentable conditions \citep{ZY} for Lasso are violated. Similarly,
Condition \ref{irepSCAD} is able to accommodate highly correlated covariates when the irrepresentable condition for the SCAD method \citep{fan2011nonconcave} fails. We illustrate this point with examples in the following subsection.

\subsection{Examples satisfying the proposed conditions}\label{examples}


In this subsection, we give some examples where the proposed irrepresentable Conditions \ref{irepLasso} and \ref{irepSCAD} still hold even when the original irrepresentable conditions for Lasso and SCAD fail, respectively. 
We suppose that $\bm{C}_n$ is a submatrix of $\bm{C}_{n+1}$ as the dimension increases.


We first consider using an extended block-exchangeable covariance matrix structure, which is 
defined as a block diagonal matrix consisting of identity matrices and $\bm{R}$:
\begin{equation}\label{block_exchange}
\bm{C}_n=diag\{\bm{I}_{q_n-q_0}, \bm{R}, \bm{I}_{p_n-q_n-(p_0-q_0)}\},
\end{equation}
where 
\begin{equation}\label{block_exchange_original}
\bm{R}_{p_0\times p_0}=\begin{pmatrix}\bm{R}^{11}&\bm{R}^{12}\\
(\bm{R}^{12})^T&\bm{R}^{22}
\end{pmatrix}
\end{equation}
is block-exchangeable with 
$$(\bm{R}^{11})_{i,j}=\left\{\begin{array}{ll}
1&i=j\\\alpha_1&i\neq j
\end{array}\right.,\quad (\bm{R}^{22})_{i,j}=\left\{\begin{array}{ll}
1&i=j\\\alpha_3&i\neq j \end{array}\right.,\quad (\bm{R}^{12})_{i,j}=\alpha_2.$$
Here, $\alpha_1$, $\alpha_2$, and $\alpha_3$ are unknown constants, $\bm{R}^{11}$ is a $q_0\times q_0$ matrix, and $p_0$ and $q_0$ are constants independent of $n$. 

The $\bm{R}$ is a fixed-dimensional and dense sub-matrix in $\bm{C}_n$. The number $q_0$ represents the number of relevant covariates whose covariance matrix $\bm{R}^{11}$ is non-sparse, while $p_0$ represents the total number of covariates whose covariance matrix $\bm{R}$ is non-sparse. There are $p_0-q_0$ irrelevant covariates whose covariance matrix $\bm{R}^{22}$ is dense. 
In addition, $\bm{R}^{12}$ represents the covariance matrix between the correlated relevant and irrelevant covariates.
We use $\bm{C}_n=diag\{\bm{I}_{q_n-q_0}, \bm{R}, \bm{I}_{p_n-q_n-(p_0-q_0)}\}$ instead of $\bm{R}$ as the covariance matrix to include a diverging and sparse  covariance matrix for high-dimensional settings. Even under the sparse covariance matrix setting, the original irrepresentable conditions still could fail. 

Similarly, we define {$\bm{C}_n=diag\{\bm{I}_{q_n-q_0}, \bm{R}_{p_0\times p_0}, \bm{I}_{p_n-q_n-(p_0-q_0)}\}$ as an extended block-autoregressive (block-AR) covariance matrix, 
where
\begin{equation}\label{AR1}
(\bm{R}^{11})_{i,j}=\alpha_1^{|i-j|}, (\bm{R}^{22})_{i,j}=\alpha_3^{|i-j|}, (\bm{R}^{12})_{i,j}=\alpha_2^{|i-(q_0+j)|}.
\end{equation}
}
{
When the covariance matrix $\bm{C}_n$ is extended block-exchangeable as in (\ref{block_exchange}), the sparsity assumption in (\ref{sparse}) holds with $K=p_0\{(p_0-1)!/\Delta_1\}^u$ and $M=p_0!/\Delta_1$, where
$(p_0-1)!$ and $p_0!$ denote the factorials of $p_0-1$ and $p_0$, respectively, and
 $\Delta_1=(1-\alpha_1)^{q_0}(1-\alpha_3)^{p_0-q_0}q_0(p_0-q_0)(\alpha_1\alpha_3-\alpha_2^2)$.
		When the covariance matrix is extended block-AR, the sparsity assumption in (\ref{sparse}) holds with $K=p_0\{(p_0-1)!/\Delta_2\}^u$ and $M=p_0!/\Delta_2$, where $\Delta_2=(1-\alpha_1^2)^{q_0}(1-\alpha_3^2)^{p_0-q_0}[1-\alpha_2^2(1-\alpha_1\alpha_2)^2(1-\alpha_3\alpha_2)^2/\{(1-\alpha_1^2)(1-\alpha_3^2)(1-\alpha_2^2)^2\}]$. Note that in both cases, the $K$ and $M$ do not depend on $p_n$. 
}
To simplify the following statements, we let $L_0=q_0/m_0$, where $m_0=|\sum_{i=q_n-q_0+1}^{q_n}\sign(\beta_i)|=|\sum_{i=q_n-q_0+1}^{q_n}\sign(\gamma_i)|>0$.
\begin{proposition}\label{exchangable1}
	Let $p_n=\exp(n^{1-2\kappa_0})$ and $q_n=n^{1/3}$ with $1/3+\tau<\kappa_0<1/2$ and $0<\tau<1/6$. 	Under the normality assumption (\ref{Normal}), suppose that $\bm{C}_n$ is an extended block-exchangeable covariance matrix of the form in (\ref{block_exchange}) with $\alpha_1, \alpha_2, \alpha_3 \in(-1,1)$ such that $\alpha_1\alpha_3\neq\alpha_2^2$ and $\bm{C}_n$ is positive definite for any large constants $q_0$ and $p_0-q_0$, where $q_0<p_0-q_0$.
	Then there exists a constant $0<\delta<1/2$ such that 
	\begin{equation}\label{OriginalIrrep}
	\|\hat{\bm{C}}_n^{21}(\hat{\bm{C}}_n^{11})^{-1}\sign\{\bm{\beta}(1)\}\|_{\infty}\ge 1 \ \ \ \ \text{ 
		with probability at least 
	} 1-O(n^{-\delta})
	\end{equation}
	if $|\alpha_2|>\alpha_1 L_0$. Conversely, (\ref{OriginalIrrep}) implies $|\alpha_2| > \alpha_1 L_0 \ge \alpha_1$, $\alpha_3\ge|\alpha_2|$, and 
	\begin{equation} \label{compareIrep}
	|\bm{V}(2)\hat{\bm{C}}_n^{21}(\hat{\bm{C}}_n^{11})^{-1}\bm{V}(1)^{-1}\sign\{\bm{\beta}(1)\}| < |\hat{\bm{C}}_n^{21}(\hat{\bm{C}}_n^{11})^{-1}\sign\{\bm{\beta}(1)\}|
	\end{equation}
	for sufficiently large constants $m_0$ and $p_0-q_0$ with probability at least $1-O(n^{-\delta})$, where the inequality holds element-wise.
\end{proposition}

\begin{proposition}\label{SCADexchangable1}
	Under the conditions of Proposition \ref{exchangable1}, if for some constant $0<\delta <1/2$, 
	\begin{equation}\label{SCADOriginalIrrep}
	\|\hat{\bm{C}}_n^{21}(\hat{\bm{C}}_n^{11})^{-1}\|_{\infty} \mathcal{P}_{\lambda}(h_{\min}) \ge 1 \ \ \ \ \text{ 
		with probability at least 
	} 1-O(n^{-\delta}),
	\end{equation}
	then $\alpha_3 \ge |\alpha_2| > \alpha_1$, and 
	\begin{equation} \label{SCADcompareIrep}
	\|\bm{V}(2)\hat{\bm{C}}^{21}_n(\hat{\bm{C}}^{11}_n)^{-1}\bm{V}(1)^{-1}\|_{\infty} \mathcal{P}_{\lambda}(h_{\min}) < \|\hat{\bm{C}}_n^{21}(\hat{\bm{C}}_n^{11})^{-1}\|_{\infty} \mathcal{P}_{\lambda}(h_{\min})
	\end{equation}
	for sufficiently large constants $m_0$ and $p_0-q_0$ with probability at least $1-O(n^{-\delta})$.
\end{proposition}

The failure of the original irrepresentable conditions (\citealp{ZY}; \citealp{fan2011nonconcave}) of the Lasso and SCAD method implies inequalities (\ref{OriginalIrrep}) and (\ref{SCADOriginalIrrep}), respectively.
By Propositions \ref{exchangable1} and \ref{SCADexchangable1}, if the original irrepresentable conditions fail,
then correlations between relevant covariates are the smallest among correlations of all covariates, followed by correlations between relevant and irrelevant covariates.
More importantly, the inequalities in (\ref{compareIrep}) and (\ref{SCADcompareIrep}) hold even when the original irrepresentable conditions are violated,
indicating that the new irrepresentable Conditions \ref{irepLasso} and \ref{irepSCAD} for the SPAC-Lasso and SPAC-SCAD, respectively, can still be valid. 

The following corollaries provide sufficient conditions for the SPAC-Lasso to be strongly sign consistent when the true covariance matrix is extended block-exchangeable as in (\ref{block_exchange}) or { extended} block-AR { with $\bm{R}$ defined} as in (\ref{AR1}). 
We also provide a similar corollary in Section S4 
in the supplementary material for the strong sign consistency of the SPAC-SCAD under the extended block-exchangeable covariance matrix structure.

\begin{corollary}
	\label{exchangeable2}
	Let $\hat{\bm{d}}$ be diagonal elements of the CLIME of $\bm{D}$. Suppose that the conditions of Proposition \ref{exchangable1} and Condition 4 
	are satisfied, and that $h_{\min}\ge n^{-\kappa_0}$. 
	If there exists a positive constant $\eta$ such that
	\begin{equation}\label{exchangeable3}
	|\alpha_2|\le (1-\eta)\left(\frac{1-\alpha_1}{1-\alpha_3}\right)^{1/2}\alpha_1 L_0,
	\end{equation}
	then the SPAC-Lasso possesses strong sign consistency and the estimator $\hat{\bm{\beta}}=\hat{\bm{V}}^{-1}\hat{\bm{\gamma}}_{Lasso}$ is consistent for sufficiently large $q_0$ and $p_0-q_0$.
	
\end{corollary}

Under the extended block-exchangeable structure with large $n$, the weak irrepresentable condition \citep{ZY} of Lasso holds for large $\alpha_1$. However, when $\alpha_1 < |\alpha_2|/L_0 \le |\alpha_2|$, the Lasso is not sign consistent by Proposition \ref{exchangable1}. In contrast, Corollary \ref{exchangeable2} shows that the SPAC-Lasso is strongly sign consistent given that $\alpha_3$ is sufficiently large, even when $\alpha_1$ is small.

\begin{corollary}
	\label{ar1}
	Suppose that conditions of Corollary \ref{exchangeable2} are satisfied, except that { $\bm{C}_n$ is an extended block-AR covariance matrix with $\bm{R}$ defined in (\ref{AR1}) and} $\alpha_1, \alpha_2, \alpha_3 \in (0,1)$, such that $2|\alpha_2-z|\{\alpha_2+1/(1+z)\}<1$, where $z=\alpha_1 \text{ or }\alpha_3$.
	{Further suppose that true coefficients of relevant covariates have the same sign.}
	If (\ref{OriginalIrrep}) is satisfied, then $\alpha_2>\alpha_1$, and the SPAC-Lasso is strongly sign consistent when there is a constant $\eta>0$ such that
	\begin{equation}\label{conditioneq}
	\max \left\{\frac{\alpha_2}{|\alpha_2-\alpha_3|}, 1\right\} \left(\frac{1-\alpha_3^2}{1-\alpha_1^2}\right)^{1/2}\frac{\alpha_2(1-\alpha_1\alpha_2)}{(1+\alpha_1)(1-\alpha_2)} \le 1-\eta.
	\end{equation}
\end{corollary}

Corollary \ref{ar1} states that, under the {extended} block-AR structure, the failure of the weak irrepresentable condition of Lasso also implies that correlations between relevant and irrelevant covariates are stronger than correlations among relevant covariates, that is, $\alpha_2>\alpha_1$. More importantly, even when the weak irrepresentable condition fails, the SPAC-Lasso is still strongly sign consistent, given that the $\alpha_3$ is sufficiently large. This is consistent with the results of the extended block-exchangeable example.

In the following proposition, we present another sufficient condition for Conditions \ref{irepLasso} and \ref{irepSCAD} of the proposed method
when the correlation structure does not have a specific form. We first introduce some notations as follows. Let $\bm{C}_n=(c_{ij})_{p\times p}$ with $c_{ij}\ge 0$, and $\bm{C}_{n,i}$ be a submatrix of $\bm{C}_n$ with the $i$th row and $i$th column removed for each $i=1, \dots, p$. Denote the $i$th column of $\bm{C}_n$ with the $i$th entry removed as $\bm{v}_{i}$, that is,  $\bm{v}_{i}=(c_{1i},\ldots,c_{i-1i},c_{i+1i},\ldots,c_{p i})^T$. In addition, we let $\varphi_i$ be the largest angle between $\bm{v}_i$ and any column vector in $\bm{C}_{n,i}$, and let $\lambda_{\min, i}$ and $\lambda_{\max, i}$ be the smallest and the largest eigenvalues of $\bm{C}_{n,i}$, respectively.
\begin{proposition}
	\label{mycondition}
	Suppose that normality assumption (\ref{Normal}), Conditions \ref{dime} and 4
	are satisfied with $\kappa_0>\max\{\kappa_2+\kappa_3, (\kappa_2+\kappa_4)/2\}$. If
	\begin{equation*}
	0\le \frac{1-\|\bm{v}_j\|_2^2/\lambda_{\max,j}}{1-\|\bm{v}_i\|_2^2 /\lambda_{\max,i} - \|\bm{v}_i\|_2^2\sin^2 \varphi_i/\lambda_{\min,i}}<g_n^2
	\end{equation*}
	holds for all $i\in \{1,\dots,q_n\}$ and $j\in \{q_n+1, \dots,p_n\}$, with $g_n=(1-\eta)/\|\bm{C}_n^{21}(\bm{C}_n^{11})^{-1}\|_{\infty}$ for some $\eta>0$, then
	$\|\bm{V}(2)\hat{\bm{C}}_n^{21}(\hat{\bm{C}}_n^{11})^{-1}\bm{V}(1)^{-1}\|_{\infty} \le 1-\eta$ with probability at least $1-O(n^{-\delta})$.
\end{proposition}


In general, when correlations between relevant and irrelevant covariates are larger than correlations among relevant covariates, the original irrepresentable conditions are likely to fail. Under this case, correlations among irrelevant covariates could be high due to the positive-definiteness constraint on correlation matrices. This indicates that, for each pair of relevant and irrelevant covariates, variables other than such a pair are more correlated with the irrelevant covariate than the relevant one. Then, the irrepresentable conditions of the proposed SPAC method are likely to hold by Proposition \ref{mycondition}. Consequently, the irrepresentable conditions for the proposed SPAC method can still be satisfied when the original irrepresentable conditions are violated.




\setcounter{equation}{0}
\section{Implementation}\label{implementation}

In this section, we discuss implementation of the proposed method with the Lasso, adaptive Lasso or SCAD penalty. For estimation of diagonal elements $\bm{d}$, 
we apply the CLIME 
in our algorithms under high-dimensional settings, which
estimates the $j$th column of the precision matrix $\bm{D}$ through the following minimization problem
\begin{eqnarray}\label{Dopt}
\underset{\bm{b}\in \mathbb{R}^{p}}{\min}|\bm{b}|_1 \ \ \  \text{                 subject to }  |\hat{\bm{C}}_n\bm{b}-\bm{e}_j|_{\infty} \le \lambda_d,
\end{eqnarray}
where $1\le j \le p$, $\bm{e}_j \in \mathbb{R}^p$ is a vector with $1$ in the $j$th coordinate and $0$ in others, $\lambda_d$ is a tuning parameter, $|\cdot|_1$ and $|\cdot|_{\infty}$ represents the $1$-norm and infinity norm of a vector, respectively. We solve the problem (\ref{Dopt}) via the  ``fastclime'' R package 
({\ttfamily https://cran.r-project.} {\ttfamily org/web/packages/fastclime/index.html}), 
and then let $\hat{d}_{jj}$ be the $j$th element of the solution.
In the fixed-dimensional settings, we use the sample precision matrix to estimate diagonal elements
\begin{equation}\label{DSP}
\hat{d}_{jj} = \{(n^{-1}\bm{X}^T\bm{X})^{-1}\}_{jj}, \  j=1,\dots, p.
\end{equation}
For the SPAC-ALasso, we estimate the initial estimator $\hat{\bm{\gamma}}_0=(\hat{\gamma}_{01}, \dots, \hat{\gamma}_{0p})^T$ in (\ref{aLassoloss}) through $\hat{\gamma}_{0j} = \hat{\beta}_{0j}/\hat{d}_{jj}^{1/2}$ $ (1\le j \le p)$, which implies that an initial estimator $\hat{\bm{\beta}}_0$ for $\bm{\beta}$ is required.
We use the ordinary least squares (OLS) estimator of $\bm{\beta}$ as the initial estimator $\hat{\bm{\beta}}_0$ under fixed-dimensional situations.
For high-dimensional settings, we first select variables via the SPAC-Lasso and then compute OLS estimators of coefficients for the selected variables. 
We let $\hat{\bm{\beta}}_0$ be the vector consisting of OLS estimators for selected covariates and zeros for non-selected covariates. 
For the tuning parameter related to the adaptive Lasso penalty in (\ref{aLassoloss}), we let $\mu=1$ in this paper and compare the proposed SPAC-ALasso to the traditional adaptive Lasso method with $\mu=1$.


We use the coordinate descent algorithm (\citealp{Fu}; \citealp{BH}) to solve the minimization problems with objective functions in (\ref{lossLasso}), (\ref{aLassoloss}), and (\ref{lossSCAD}) for the SPAC-Lasso, SPAC-ALasso, and SPAC-SCAD, respectively.
We illustrate this with $p=1$ first. The unpenalized least squares solution of the univariate setting is $z=\bm{X}^T\bm{y}/(n\hat{d}^{1/2})$. 
Accordingly, the proposed SPAC-Lasso, SPAC-ALasso, and SPAC-SCAD estimators have closed forms
\begin{equation}\label{fLasso}
\hat{\gamma}_{Lasso}(z, \lambda) = \sign(z)(|z|-\lambda)_+, \ \ \ \hat{\gamma}_{ALasso}(z, \lambda, \hat{\gamma}_0) = \sign(z)\left(|z|-\frac{\lambda}{|\hat{\gamma}_0|}\right)_+,
\vspace{-3mm}
\end{equation}
\begin{equation}\label{fSCAD}
\hat{\gamma}_{SCAD}(z, \lambda, a) = \left\{\begin{array}{ll}
\sign(z) (|z|-\lambda)_+&\text{if } |z|\le 2\lambda\\
\{(a-1)z-\sign(z)a\lambda\}/(a-2)&\text{if } 2\lambda<|z|\le a \lambda\\
z&\text{if } |z|>a \lambda,
\end{array}\right.
\end{equation}
respectively.


For a multivariate case, we use these univariate solutions to obtain coordinate-wise minimizers, except that we replace $z$ in (\ref{fLasso}) and (\ref{fSCAD}) by the unpenalized solution of the regression with the partial residual of $\bm{x}_j$ ($1\le j \le p$) as response
\begin{equation}\label{zj}
z_j=\bm{x}_j^T\bm{r}_{-j}/(n\hat{d}_{jj}^{1/2})=\bm{x}_j^T\bm{r}/(n\hat{d}_{jj}^{1/2})+\gamma_j^{(l-1)},
\end{equation}
where $\bm{r}_{-j}=\bm{y}-\sum_{i\neq j} \bm{x}_i\hat{d}_{ii}^{1/2}\gamma_i^*$ is $\bm{x}_j$'s partial residual, $\bm{r}=\bm{y}-\sum_{i=1}^{p}\bm{x}_i\hat{d}_{ii}^{1/2}\gamma_i^*$, and $\bm{\gamma^*}=(\gamma_1^*,\dots, \gamma_{p}^*)^T$ is the most recent updated estimator for $\bm{\gamma}$. A complete algorithm for the SPAC-SCAD is provided in Algorithm 1, including estimation of the 
$\bm{d}$ in Step 2 and the coordinate descent method in Step 4. Algorithms of the SPAC-Lasso and SPAC-ALasso are similar to Algorithm 1, except that we replace $\hat{\gamma}_{SCAD}$ in Step 4 by $\hat{\gamma}_{Lasso}$ or $\hat{\gamma}_{ALasso}$ in (\ref{fLasso}), respectively. 

\begin{table}[H]\centering
	\renewcommand{\arraystretch}{0.8}
	\begin{tabular}{c}
		\hline
		\textbf{Algorithm 1} (SPAC-SCAD) 
		\\
		\hline
		\parbox{14cm}{
			\begin{tabbing}
				\enspace 1. Set $l=1$. Set tolerance $\epsilon$, initial values $\bm{\gamma}^{(0)}$, and tuning parameters $\lambda$ and $a$.\\
				\enspace 2. Calculate $\hat{\bm{d}}$ using (\ref{DSP}) or through solving (\ref{Dopt}) for $j=1, \dots, p$.\\
				\enspace 3. Calculate $\bm{r}^{(0)}=\bm{y}-\sum_{j=1}^{p}\bm{x}_j\hat{d}_{jj}^{1/2}\gamma_j^{(0)}$.\\
				\enspace 4. For $j=1,\dots,p,$ estimate $\gamma^{(l)}_j$ as follows\\
				\qquad Calculate $z_j$ using (\ref{zj});\\
				\qquad Calculate $\gamma^{(l)}_j = \hat{\gamma}_{SCAD} (z_j, \lambda, a)$ using (\ref{fSCAD});\\
				\qquad Update $\bm{r}^{(l)}=\bm{r}^{(l-1)}-\bm{x}_j\hat{d}_{jj}^{1/2}(\gamma_j^{(l)}-\gamma_j^{(l-1)})$.\\
				\enspace 5. Iterate Step 4 until a convergence criterion is satisfied, for example, \\ \hspace{6.5mm}$\underset{j}{\text{min}}\left\{\left|(\gamma_j^{(l)}-\gamma_j^{(l-1)})/\gamma_j^{(l-1)}\right|\right\}<\epsilon$.
			\end{tabbing}
		}\\
		\hline
	\end{tabular}
\end{table}

\setcounter{equation}{0}
\section{Simulations} \label{simulation}

In this section, we compare the performance of the proposed method and existing model selection approaches in simulation studies. We generate data $100$ times based on a linear regression model, $\bm{y}=\bm{X}\bm{\beta}+N_n(\bm{0}, \bm{I}_n)$, 
where $\bm{X}$ is a $n\times p$ matrix and $\bm{\beta}$ is a $p\times 1$ vector. Each row of the design matrix $\bm{X}$ is independently and identically distributed from a multivariate normal distribution with mean $\bm{0}_{p\times 1}$ and a block-exchangeable covariance matrix $\bm{C}_{p\times p}$ of the form in (\ref{block_exchange_original}) with the parameters $\bm{\alpha}=(\alpha_1,\alpha_2,\alpha_3)^T$.
The first $q$ elements in the coefficient vector $\bm{\beta}$ are 
nonzero and take value 
$\beta_s$, while the remaining ones are zero. 

We implement the Lasso, adaptive Lasso, and SCAD methods via the coordinate descent algorithm (\citealp{Fu}; \citealp{BH}). Since the purpose of the proposed method is to provide model selection consistency when the traditional methods fail, we first check whether the original weak irrepresentable condition is satisfied or not for the covariates selected by Lasso. If the weak irrepresentable condition is violated, we adopt the proposed method; otherwise, the standard Lasso, adaptive Lasso, and SCAD methods can still be applicable. 
We use the ``pcalg'' R package 
({\ttfamily https://cran.r-project.org/web/packages/\linebreak pcalg/index.html})
to implement the PC-simple algorithm  
with a significance level $0.05$, which is a method based on partial correlations.
The Farm-Select method 
is implemented via the ``FarmSelect'' R package 
({\ttfamily https://cran.r-project.org/web/packages/FarmSelect/\linebreak index.html}). 
In each penalty-based method, the tuning parameter $\lambda$ is selected by the extended BIC (EBIC) which is effective for small $n$ but large $p$ \citep{chen2008extended}. For the SCAD method and the proposed SPAC-SCAD method, we choose $a=3.7$ \citep{FL}. For the adaptive Lasso, we apply the Lasso estimator as the initial estimator for weighting. 

To evaluate the performance of each method, we compute the false negative rate (FNR) and false positive rate (FPR) as follows
\[
\frac{\sum_{j=1}^{p} I(\hat{\beta}_j = 0, \beta_j \neq 0)}{\sum_{j=1}^{p} I(\beta_j \neq 0)}, \ \ \ 
\frac{\sum_{j=1}^{p} I(\hat{\beta}_j \neq 0, \beta_j = 0)}{\sum_{j=1}^{p} I(\beta_j = 0)},
\]
respectively, where $I(\cdot)$ is an indicator function. The FNR represents the proportion of relevant covariates which are not selected, while the FPR represents the proportion of selected irrelevant covariates. We define the overall false rate of a method as the summation of the FNR and FPR. A method with smaller overall false rate indicates a better performance in model selection. 
We calculate means of the FNRs and FPRs for all the implemented methods using $100$ replications.

\textbf{Setting $1$:} 
Let $p=250$, $q=5$, $n=80$, $\beta_s=0.4$, and $\bm{\alpha}=(0.1, 0.3, 0.8)^T$, $(0.2, 0.4, 0.8)^T$,  $(0.3, 0.5, $ $0.8)^T$, or $(0.5,0.7,0.9)^T$.


Table \ref{TS1} shows that the proposed method performs better than existing model selection approaches under Setting $1$. Specifically, the ratio of overall false rate of each penalty-based method to that of the proposed method with the same penalty function is greater than $1$ across all covariance matrices. Also, the overall false rates of the Farm-Select method and the PC-simple algorithm are both larger than that of the proposed SPAC-SCAD.
In particular, the ratio of overall false rates is the largest when $\bm{\alpha}=(0.5, 0.7, 0.9)^T$ where covariates are most correlated. For example, the ratio between the traditional Lasso and the proposed SPAC-Lasso is $6.637$ when $\bm{\alpha}=(0.5, 0.7, 0.9)^T$, which is much larger than corresponding ratios under other $\bm{\alpha}$'s. 

Moreover, the FNRs of the SPAC-Lasso, SPAC-ALasso, and SPAC-SCAD are smaller than those of the traditional Lasso, adaptive Lasso, and SCAD methods given each $\bm{\alpha}$, respectively. This also holds for the FPR. In addition, we present violation rates in the last row of Table \ref{TS1}, which is the percentage of the original weak irrepresentable condition being violated based on 100 simulated data. The violation rates are all close to $1$ since the original weak irrepresentable condition does not hold for the true covariance matrices in this setting.

\textbf{Setting $2$:} Let $p=1000$, $q=20$, $n=150$, $\bm{\alpha}=(0.3,0.5,0.8)^T$, and $\beta_s=0.2, 0.3, 0.4, 0.5$, or $0.6$.

\begin{table}\small\centering
	\renewcommand{\arraystretch}{0.65}
	\caption{\small \linespread{1.3}\selectfont{} Results of Setting $1$. The ``Ratio'' for each penalty-based approach is the ratio of FPR$+$FNR calculated from the traditional method to the FPR$+$FNR from the proposed method with the same penalty. The ``Ratio'' for Farm-Select (or PC-simple) is the ratio of FPR$+$FNR for Farm-Select (or PC-simple) to that of SPAC-SCAD. ``Violate'' represents the percentage of the original weak irrepresentable condition being violated based on selection results of Lasso in 100 simulated data.}
	\label{TS1}\par
	\vskip .2cm
	\begin{tabular}{cccccc}
		\Xhline{2\arrayrulewidth}
		\multicolumn{2}{c}{$\bm{\alpha}$}&(0.1, 0.3, 0.8)&(0.2, 0.4, 0.8)&(0.3, 0.5, 0.8)&(0.5, 0.7, 0.9)\\
		\Xhline{2\arrayrulewidth}
		& FNR &0.804 &0.718 &0.744 &0.744\\
		\raisebox{2ex}[0pt]{Lasso} 	& FPR &0.002&0.005&0.007&0.018\\
		\hline
		& FNR &0.510 &0.382 &0.460 &0.112\\
		& FPR &0.002 &0.003 &0.006 &0.003\\
		\raisebox{4ex}[0pt]{SPAC-Lasso} & \textbf{Ratio} &1.576 &1.876 &1.614 &\textbf{6.637}\\
		\Xhline{2\arrayrulewidth}
		& FNR &0.794&0.778&0.794&0.890\\
		\raisebox{2ex}[0pt]{ALasso} 	& FPR &0.001&0.001&0.003&0.006\\
		\hline
		& FNR &0.500&0.430&0.528&0.384\\
		& FPR &0.000&0.001&0.002&0.002\\
		\raisebox{4ex}[0pt]{SPAC-ALasso} & \textbf{Ratio} &1.589 &1.808 &1.505 &\textbf{2.321}\\
		\Xhline{2\arrayrulewidth}
		& FNR &0.148&0.196&0.380&0.859\\
		\raisebox{2ex}[0pt]{SCAD} 	& FPR &0.126&0.093&0.057&0.004\\
		\hline
		& FNR &0.126&0.120&0.214&0.303\\
		& FPR &0.052&0.043&0.038&0.003\\
		\raisebox{4ex}[0pt]{SPAC-SCAD} & \textbf{Ratio} &1.542 &1.775 &1.734 &\textbf{2.821}\\
		\Xhline{2\arrayrulewidth}
		& FNR &0.200&0.600&0.200&1.000\\
		& FPR &0.065&0.029&0.065&0.004\\
		\raisebox{4ex}[0pt]{Farm-Select} & \textbf{Ratio} &1.489&\textbf{3.859}&1.052&\textbf{3.281}\\
		\Xhline{2\arrayrulewidth}
		& FNR &0.496&0.530&0.696&0.892\\
		& FPR &0.003&0.004&0.005&0.007\\
		\raisebox{4ex}[0pt]{PC-simple} & \textbf{Ratio} &\textbf{2.809} &\textbf{3.273} &\textbf{2.782} &\textbf{2.937}\\
		\Xhline{2\arrayrulewidth}
		\multicolumn{2}{c}{Violate}&0.900&0.940&0.860&0.970\\
		\Xhline{2\arrayrulewidth}
	\end{tabular}
\end{table}

We consider high-dimensional situations with one thousand covariates in Setting $2$. Results in Table \ref{S23} show that the proposed method still outperforms other competing methods in terms of overall false rate. In addition, the ratios between overall false rates are larger for scenarios with larger $\beta_s$, indicating that the proposed method improves existing methods more when signals are stronger. The FNR and FPR of the PC-simple algorithm for relatively larger $\beta_s$ are not provided in Table \ref{S23} since the PC-simple algorithm is quite time-consuming under  settings with strong signals and thousands of correlated potential predictors. It takes more than a few hours to run the algorithm for only one replication. However, we can still observe that the proposed SPAC-SCAD outperforms the PC-simple algorithm based on results under $\beta_s=0.2$ and $\beta_s=0.3$.

We incorporate binary covariates in the following Setting $3$. We first simulate data from a multivariate normal distribution with mean $\bm{0}$ and covariance matrix $\bm{C}$ of the form in  (\ref{block_exchange_original}), and then transform $2$ relevant and $60$ irrelevant covariates $X_j$ to $\sign(X_j)$.

\textbf{Setting $3$:} Let $p=250$, $q=5$, $n=80$,  $\bm{\alpha}=(0.5,0.7,0.9)^T$, and $\beta_s=0.2, 0.3, 0.4, 0.5, 0.6$, or $0.7$. 


The proposed method also performs better than other methods even when we have binary potential predictors, according to the results in Table \ref{S23}. For instance, when $\beta_s=0.5$, the FNR of the SPAC-ALasso is $0.320$, only $43.4\%$ of the FNR of the adaptive Lasso method, indicating that the proposed SPAC-ALasso selects more relevant covariates. Similarly, the FPR of the SPAC-ALasso is smaller than that of the adaptive Lasso, implying that the proposed SPAC-ALasso selects fewer irrelevant covariates. In addition, the overall false rate of the SPAC-SCAD decreases much faster than that of the PC-simple algorithm as $\beta_s$ increases, which is consistent with the fact that partial correlation is unable to fully use signal strength due to its bounded range.


\begin{table}[H]\centering
	\renewcommand{\arraystretch}{0.65}
		\caption{	\linespread{1.3}\selectfont{}
		\small Results of Settings $2$ and $3$. The ``Ratio'' for each penalty-based approach is the ratio of FPR$+$FNR calculated from the traditional method to the FPR$+$FNR from the proposed method with the same penalty. The ``Ratio'' for Farm-Select (or PC-simple) is the ratio of FPR$+$FNR for Farm-Select (or PC-simple) to that of SPAC-SCAD. ``Violate'' represents the percentage of the original weak irrepresentable condition being violated based on selection results of Lasso in 100 simulated data.}
	\label{S23}\par
	\vskip .2cm
	\resizebox{\textwidth}{!}{%
		\begin{tabular}{>{\centering}p{20mm}c|ccccc|cccccc}
			\Xhline{2\arrayrulewidth}
			\multicolumn{2}{c|}{Setting}&\multicolumn{5}{c|}{Setting $2$}&\multicolumn{6}{c}{Setting $3$}\\
			\hline
			\multicolumn{2}{c|}{$\beta_s$}&0.2&0.3&0.4&0.5&0.6	&0.2&0.3&0.4&0.5&0.6&0.7\\
			\Xhline{2\arrayrulewidth}
			& FNR &0.986&0.972&0.852&0.586&0.415	&0.976 &0.912 &0.782 &0.642 &0.388 &0.180\\
			\raisebox{2ex}[0pt]{Lasso} 	& FPR	&0.007&0.008&0.010&0.016&0.019	&0.003 &0.009 &0.016 &0.018 &0.025 &0.028\\
			\hline
			& FNR	&0.921&0.639&0.357&0.094&0.020	&0.904 &0.692 &0.334 &0.182 &0.044 &0.006\\
			& FPR 	&0.003&0.007&0.012&0.019&0.021	&0.002 &0.003 &0.009 &0.011 &0.013 &0.014\\
			\raisebox{4ex}[0pt]{SPAC-Lasso} & \textbf{Ratio}	&1.075&1.516&\textbf{2.334}  &\textbf{5.328} &\textbf{10.625}	&1.082  &1.325  &\textbf{2.329}  &\textbf{3.425}  &\textbf{7.214} &\textbf{10.557}\\
			\Xhline{2\arrayrulewidth}
			& FNR 	&0.998&0.990&0.936&0.659&0.466	&0.982 &0.954 &0.854 &0.738 &0.496 &0.300\\
			\raisebox{2ex}[0pt]{ALasso} 	& FPR 	&0.002&0.002&0.002&0.003&0.002	& 0.002 &0.004 &0.005 &0.006 &0.005 &0.005\\
			\hline
			& FNR 	&0.969&0.806&0.565&0.340&0.249	&0.914 &0.744 &0.478 &0.320 &0.108 &0.042\\
			& FPR 	&0.001&0.001&0.001&0.001&0.000	&0.001 &0.002 &0.002 &0.002 &0.002 &0.001\\
			\raisebox{4ex}[0pt]{SPAC-ALasso} & \textbf{Ratio} 	&1.031 &1.229 &1.657 &1.939 &1.881	&1.075 &1.284 &1.788 &\textbf{2.309} &\textbf{4.569} &\textbf{7.080}\\
			\Xhline{2\arrayrulewidth}
			& FNR &0.827&0.702&0.580&0.323&0.073	&0.824 &0.830 &0.780 &0.612 &0.476 &0.242\\
			\raisebox{2ex}[0pt]{SCAD} 	& FPR & 0.005& 0.008& 0.008& 0.007&0.004	&0.024 &0.008 &0.008 &0.009 &0.006 &0.005\\
			\hline
			& FNR &0.513&0.296& 0.108&0.013&0.001	&0.774 &0.499 &0.298 &0.158 &0.078 &0.032\\
			& FPR &0.007&0.007&0.004&0.002&0.001	&0.005 &0.006 &0.007 &0.004 &0.004 &0.002\\
			\raisebox{4ex}[0pt]{SPAC-SCAD} & \textbf{Ratio} &1.599  &\textbf{2.339}  &\textbf{5.235} &\textbf{22.118} &\textbf{44.648}	&1.090 &1.659 &\textbf{2.585} &\textbf{3.825} &\textbf{5.908} &\textbf{7.275}\\
			\Xhline{2\arrayrulewidth}
			& FNR &1.000 &1.000 &1.000 &0.850 &0.800	&1.000 &1.000 &1.000 &1.000 &0.400 &0.200 \\
			& FPR &0.000 &0.001 &0.005 &0.006 &0.006	&0.016 &0.016 &0.024 &0.000 &0.029 &0.029\\
			\raisebox{4ex}[0pt]{Farm-Select} & \textbf{Ratio} &1.923   &\textbf{3.296}   &\textbf{8.944}  &\textbf{$>$50} &\textbf{$>$50}	&1.306 &\textbf{2.011} &\textbf{3.359} &\textbf{6.154} &\textbf{5.250} &\textbf{6.712}\\
			\Xhline{2\arrayrulewidth}
			& FNR &0.992&0.997&---&---&---	&0.918 &0.876 &0.844 &0.786 &0.726 &0.682\\
			& FPR &0.005&0.007&---&---&---	&0.005 &0.006 &0.007 &0.007 &0.007 &0.007\\
			\raisebox{4ex}[0pt]{PC-simple} & \textbf{Ratio} &1.917  &\textbf{3.305}&---&---&---	&1.186  &1.746  &\textbf{2.789} &\textbf{4.880}  &\textbf{8.981} &\textbf{20.242}\\
			\Xhline{2\arrayrulewidth}
			\multicolumn{2}{c}{Violate}&0.980 &0.970 &1.000 &1.000 &1.000	&0.909 &0.899 &0.980 &0.960 &0.970 &1.000\\
			\Xhline{2\arrayrulewidth}
	\end{tabular}}
\end{table}

Since the estimation of diagonal elements $\bm{d}$ could be inaccurate, we investigate the robustness of the proposed method with respect to the estimation of $\bm{d}$ in the following Setting $4$. In this setting, we replace $\hat{d}_{jj}$ in the implementation of SPAC-Lasso by $\hat{d}_{jj}+u_j$ for each $j=1, \dots, p$, where $u_j$'s are independently and identically distributed from a truncated normal distribution with minimum value $\max_{1\le j \le p}\{-\hat{d}_{jj}\}$, mean $0$, and variance $\sigma_u^2$. Here, we require the random noise $u_j\ge \max_{1\le j \le p}\{-\hat{d}_{jj}\}$ to ensure that $\hat{d}_{jj}+u_j$ is positive for each $j=1, \dots, p$.

\textbf{Setting $4$:} 
Let $p=500$, $q=6$, $n=100$, $\beta_s=0.3$, and $\bm{\alpha}=(0.1, 0.3, 0.8)^T$ or $(0.2,0.4,0.8)^T$.  The variance parameter $\sigma_u=0, 1, 3,$ or $5$.

The results of Setting $4$ in Table \ref{S4} show that the overall false rate of the SPAC-Lasso with noise  increases as the variance $\sigma_u^2$ increases, but this overall false rate is still smaller than that of the Lasso method. For example, when $\bm{\alpha}=(0.1, 0.3, 0.8)^T$, the overall false rate of the SPAC-Lasso with $\sigma_u=5$ is $0.919$ larger than that with $\sigma_u=0$, but smaller than $0.989$ which is the overall false rate of the Lasso method. Thus, the proposed method is robust to certain errors in the estimation of $\bm{d}$. Although we use the CLIME for the estimation of $d$ in this paper, this can also be replaced with other consistent estimators.

\begin{table}\small\centering
	\renewcommand{\arraystretch}{0.65}
		\caption{Results of Setting $4$.}
	\label{S4}\par
	\vskip .2cm
	\begin{tabular}{c|c|cccc|c|cccc}
		\Xhline{2\arrayrulewidth}
		$\bm{\alpha}$&\multicolumn{5}{c|}{(0.1, 0.3, 0.8)}&\multicolumn{5}{c}{(0.2, 0.4, 0.8)}\\
		\hline
		Method&Lasso&\multicolumn{4}{c|}{SPAC-Lasso}&Lasso&\multicolumn{4}{c}{SPAC-Lasso}\\
		\hline
		$\sigma_u$&---&0&1&3&5	&---&0&1&3&5\\
		\hline
		FNR &0.988&0.843&0.903&0.895&0.918	&0.958&0.848 &0.867 &0.915&0.908\\
		FPR	&0.001&0.000 &0.000&0.001&0.001	&0.001 &0.001 &0.001 &0.002&0.002\\
		\textbf{FNR$+$FPR}	&0.989&0.843  &0.903 &0.896&0.919	&0.960  &0.849   &0.868 &0.917 &0.910\\
		\Xhline{2\arrayrulewidth}
	\end{tabular}
\end{table}

\textbf{Setting $5$:} Let $p=150$, $q=3$, $n=80$, and $\beta_s=0.5$. The parameters $\bm{\alpha}=(\alpha_1,\alpha_2,\alpha_3)^T$ are $(0.2, 0.4, 0.8)$, $(0.8, 0.4, 0.2)$, $(0.1, 0.3, 0.8)$, $(0.8, 0.3, 0.1)$, $(0.2, 0.4, 0.7)$, $(0.7, 0.4, 0.2)$, $(0.4, 0.5, \linebreak 0.7)$, or $(0.7, 0.5, 0.4)$.

In Setting $5$, we examine the robustness of the proposed method when the original weak irrepresentable condition holds. As shown in Tables \ref{setting5_1}--\ref{setting5_2}, the proposed method still outperforms the existing methods in terms of FNR$+$FNP when $\alpha_3>\alpha_1$. In addition, the proposed method performs comparably to the existing methods when $\alpha_1>\alpha_3$, where the original weak irrepresentable condition holds. For example, the ratios of overall false rates of the adaptive Lasso method to those of the proposed SPAC-ALasso are greater than $1.5$ when $\alpha_3>\alpha_1$, and are equal to or quite close to $1$ when $\alpha_1>\alpha_3$.
In summary, the proposed method performs similarly to the regular penalization method when the weak irrepresentable condition holds, but performs much better than the existing method when the condition fails.

		\begin{table}[H]\small\centering 
	\renewcommand{\arraystretch}{0.65}
	\caption{\linespread{1.3}\selectfont{} \small 
 Results of Setting $5$. The ``Ratio'' for each penalty-based approach is the ratio of FPR$+$FNR calculated from the traditional method to the FPR$+$FNR from the proposed method with the same penalty. 
	``Violate'' represents the percentage of the original weak irrepresentable condition being violated based on selection results of Lasso in 100 simulated data.}\label{setting5_1}
\par
	\vskip .2cm
	\begin{tabular}{cccccc}
		\Xhline{2\arrayrulewidth}
		\multicolumn{2}{c}{$\bm{\alpha}$}&(0.2, 0.4, 0.8)&(0.8, 0.4, 0.2)&(0.1, 0.3, 0.8)&(0.8, 0.3, 0.1)\\
		\Xhline{2\arrayrulewidth}
		& FNR 
		& 0.240 & 0.110 & 0.357 & 0.103\\
		\raisebox{2ex}[0pt]{Lasso} 	& FPR 
		& 0.009 & 0.002 & 0.006 & 0.002\\
		\hline
		& FNR 
		& 0.143 & 0.123 & 0.123 & 0.110\\
		& FPR 
		& 0.003 & 0.002 & 0.003 & 0.002\\
		\raisebox{4ex}[0pt]{SPAC-Lasso} & \textbf{Ratio} 
		& 1.699 & 0.895 & 2.871 & 0.942\\
		\Xhline{2\arrayrulewidth}
		& FNR 
		& 0.297 & 0.393 &0.257 & 0.373\\
		\raisebox{2ex}[0pt]{ALasso} 	& FPR 
		& 0.002 & 0.001 & 0.001 & 0.001\\
		\hline
		& FNR 
		& 0.160 & 0.413 & 0.147 & 0.373\\
		& FPR 
		& 0.001 & 0.001 & 0.001 & 0.001\\
		\raisebox{4ex}[0pt]{SPAC-ALasso} & \textbf{Ratio} 
		& 1.852 & 0.952 & 1.748 & 1.000\\
		\Xhline{2\arrayrulewidth}
		& FNR 
		& 0.073 & 0.647 & 0.057 & 0.660\\
		\raisebox{2ex}[0pt]{SCAD} 	& FPR 
		& 0.012 & 0.002 & 0.009 & 0.002\\
		\hline
		& FNR 
		& 0.050 & 0.657 & 0.030 & 0.663\\
		& FPR 
		& 0.007 & 0.002 & 0.006 & 0.002\\
		\raisebox{4ex}[0pt]{SPAC-SCAD} & \textbf{Ratio} 
		& 1.493 & 0.985 & 1.830 & 0.995\\
		\Xhline{2\arrayrulewidth}
		\multicolumn{2}{c}{Violate}
		& 0.797 & 0.037 & 0.743 & 0.007\\
		\Xhline{2\arrayrulewidth}
	\end{tabular}
\end{table}

		\begin{table}[H]\small\centering  
	\renewcommand{\arraystretch}{0.65}
	\caption{\linespread{1.3}\selectfont{} \small 
	\selectfont{} Results of Setting $5$. The ``Ratio'' for each penalty-based approach is the ratio of FPR$+$FNR calculated from the traditional method to the FPR$+$FNR from the proposed method with the same penalty. 
	``Violate'' represents the percentage of the original weak irrepresentable condition being violated based on selection results of Lasso in 100 simulated data.} \label{setting5_2}
\par
	\vskip .2cm
	\begin{tabular}{cccccc}
		\Xhline{2\arrayrulewidth}
		\multicolumn{2}{c}{$\bm{\alpha}$}&(0.2, 0.4, 0.7)&(0.7, 0.4, 0.2)&(0.4, 0.5, 0.7)&(0.7, 0.5, 0.4)\\
		\Xhline{2\arrayrulewidth}
		& FNR 
		& 0.520 & 0.103 & 0.353 & 0.170\\
		\raisebox{2ex}[0pt]{Lasso} 	& FPR 
		& 0.010 & 0.004 & 0.013 & 0.004\\
		\hline
		& FNR 
		& 0.303 & 0.110 & 0.163 & 0.193\\
		& FPR 
		& 0.007 & 0.004 & 0.006 & 0.004\\
		\raisebox{4ex}[0pt]{SPAC-Lasso} & \textbf{Ratio} 
		& 1.708 & 0.943 & 2.158 & 0.884\\
		\Xhline{2\arrayrulewidth}
		& FNR 
		& 0.593 & 0.310 & 0.377 & 0.327\\
		\raisebox{2ex}[0pt]{ALasso} 	& FPR 
		& 0.002 & 0.002 & 0.003 & 0.002\\
		\hline
		& FNR 
		& 0.380 & 0.310 & 0.200 & 0.350\\
		& FPR 
		& 0.001 & 0.002 & 0.002 & 0.003\\
		\raisebox{4ex}[0pt]{SPAC-ALasso} & \textbf{Ratio} 
		& 1.562 & 1.000 & 1.882 & 0.931\\
		\Xhline{2\arrayrulewidth}
		& FNR 
		& 0.280 & 0.463 & 0.187 & 0.493\\
		\raisebox{2ex}[0pt]{SCAD} 	& FPR 
		& 0.012 & 0.008 & 0.015 & 0.009\\
		\hline
		& FNR 
		& 0.140 & 0.473 & 0.133 & 0.497\\
		& FPR 
		& 0.006 & 0.008 & 0.010 & 0.010\\
		\raisebox{4ex}[0pt]{SPAC-SCAD} & \textbf{Ratio} 
		& 1.991 & 0.980 & 1.409 & 0.991\\
		\Xhline{2\arrayrulewidth}
		\multicolumn{2}{c}{Violate}
		& 0.883 & 0.007 & 0.890 & 0.133\\
		\Xhline{2\arrayrulewidth}
	\end{tabular}
\end{table}

\setcounter{equation}{0}
\section{Real data application}\label{real}

In this section, we apply the proposed method to high-dimensional genetic data collected in the Detroit neighborhood health study
(\verb+https://dnhs.unc.edu/+),
which is a representative study focusing on post-traumatic stress disorder (PTSD) 
of African American adults in Detroit, Michigan. This study collects gene expression data and post-traumatic checklists based on incident trauma exposures, which is a $17$-item self-reported measures of PTSD symptoms. We treat the average of the $17$ post-traumatic checklist scores as the
response $Y$. Studies (\citealp{logue2015analysis}; \citealp{kuan2017gene}) show that gene expression is associated with PTSD. To identify gene probes which are relevant to PTSD, we consider using all the gene probes as potential predictors.

Since the number of all gene probes is over $15,000$ and the sample size is only $93$, we first apply screening to the gene probes based on correlations among probes and marginal correlations between probes and $Y$. For each probe ${X}_j$, we let $\bm{c}_j$ denote the vector consisting of correlations between this probe and other probes. Since the proposed method targets correlated data, we consider ${X}_j$ to be correlated with others and select it if the average absolute value of elements in $\bm{c}_j$ is greater than $0.1$.
Moreover, we calculate marginal correlations between selected probes and the response variable, and filter out probes with absolute values of the marginal correlations less than $0.15$, which are unlikely to be important probes. After the screening, we retain $3591$ gene probes for further analysis.


To evaluate the performance of different methods, we randomly partition all the observations into $95\%$ for training and $5\%$ for testing $100$ times. For each method, we estimate parameters using the training sets, calculate the mean number of selected probes, and compute the average of prediction mean squared errors (PMSE) in testing sets based on $100$ replications.
However, the PMSEs of the PC-simple algorithm and the Farm-Select method are unavailable, since the PC-simple algorithm only provides variable selection results without coefficient estimation and the R package of the Farm-Select method does not have an intercept in the model. To calculate prediction errors for the two methods and compare them to other methods, we adopt the OLS to estimate the coefficients of probes selected by each method and calculate PMSE based on the OLS estimation, denoted by OLS-PMSE.
The original weak irrepresentable condition fails in each training set based on the selection results of Lasso, indicating that the proposed method is more suitable for the data than traditional methods.

Table \ref{Tfive} provides averages of PMSEs, OLS-PMSEs, and numbers of selected probes, respectively, for all the methods. 
According to the table, the proposed method produces smaller PMSE and smaller OLS-PMSE than existing methods.
In particular, the average OLS-PMSE of Lasso is $18.7\%$ more than that of the SPAC-Lasso. Similarly, the average PMSE of the traditional adaptive Lasso and SCAD methods are $16.2\%$ and $17.3\%$ more than those of the proposed SPAC-ALasso and SPAC-SCAD,
respectively.
Moreover, in terms of OLS-PMSE, the Farm-Select method and PC-simple algorithm perform worse than the proposed method. Among all the methods, the SPAC-ALasso produces smaller PMSE with relatively fewer selected probes, while the prediction errors of methods with the SCAD penalty are larger than errors of methods with other penalties.

\begin{table}[H]\centering
	\renewcommand{\arraystretch}{0.65}
	\caption{Average results for the real data.}
	\label{Tfive}\par
	\vskip .2cm
	\begin{tabular}{c|ccc}
		\hline
		& PMSE & OLS-PMSE & NS\\
		\hline
		Lasso			&0.9306&0.9868&73\\
		\textbf{SPAC-Lasso}&\textbf{0.8283}&\textbf{0.8310}&74\\
		ALasso			&0.9568&1.0406&20\\
		\textbf{SPAC-ALasso}&\textbf{0.8232}&\textbf{0.9101}&22\\
		SCAD			&1.3353&1.3164&38\\
		\textbf{SPAC-SCAD}  &\textbf{1.1387}&\textbf{1.1298}&39\\
		Farm-Select	  &---&1.2429&40\\
		PC-simple	   &---&1.3278&5\\
		\hline
	\end{tabular}
\end{table}

In addition, we apply these methods to all the samples and summarize selected probes in 
tables in Section S3 
of the supplementary material. 
On one hand, {\fontfamily{qhv}\selectfont \small \textit{ILMN\_1716728}}, {\fontfamily{qhv}\selectfont \small \textit{ILMN\_1682259}}, {\fontfamily{qhv}\selectfont \small \textit{ILMN\_3307729}}, {\fontfamily{qhv}\selectfont \small \textit{ILMN\_1670134}},  {\fontfamily{qhv}\selectfont\small \textit{ILMN\_1793201}},  {\fontfamily{qhv}\selectfont\small \textit{ILMN\_1811507}}, {\fontfamily{qhv}\selectfont \small \textit{ILMN\_1656111}}, and {\fontfamily{qhv}\selectfont\small \textit{ILMN\_3248844}} are common probes selected by the Lasso, SPAC-Lasso, ALasso, SPAC-ALasso, SCAD, SPAC-SCAD, and Farm-Select.
Thus, these probes are very likely to be associated with the response. Among them, {\fontfamily{qhv}\selectfont\small \textit{ILMN\_1716728}}, {\fontfamily{qhv}\selectfont\small \textit{ILMN\_3307729}}, and {\fontfamily{qhv}\selectfont\small \textit{ILMN\_3248844}} are also selected by the PC-simple algorithm, indicating that these three probes are extremely likely to be relevant to PTSD. On the other hand, {\fontfamily{qhv}\selectfont\small\textit{ILMN\_1663035}} from the \textit{SREBF1} gene 
is only selected by the proposed SPAC-Lasso and SPAC-ALasso. 
According to the existing literature \citep{kuan2017epigenome}, the \textit{SREBF1} gene is indeed associated with PTSD.

In conclusion, the proposed method leads to smaller PMSE and OLS-PMSE than existing variable selection methods with similar numbers of selected probes, showing that the proposed SPAC strategy improves the accuracy of variable selection.

\setcounter{equation}{0}
\section{Discussion}

We propose a new variable selection approach to address the problem where the original irrepresentable conditions fail due to the strong dependency between relevant and irrelevant covariates. The violation of the irrepresentable conditions leads to inconsistency of model selection based on traditional methods. In this paper, we introduce a semi-standard partial covariance (SPAC), which has a clear geometric interpretation based on projections and takes advantage of both coefficients $\bm{\beta}$ and partial correlations. Moreover, we develop a SPAC method which penalizes SPACs instead of coefficients $\bm{\beta}$ or partial correlations alone to mitigate selection of irrelevant covariates that are strongly correlated with relevant covariates.

We establish the strong sign consistency of the proposed SPAC-Lasso and SPAC-SCAD
under  high-dimensionality. 
Specifically, we transform irrepresentable conditions to achieve variable selection consistency, which solves the problem when the Lasso or SCAD method is not sign consistent. Since our goal is to target situations where the traditional methods fail, we first check whether the original weak irrepresentable condition holds or not. If it is violated, numerical studies show that the proposed approach is more effective, and outperforms the traditional variable selection methods. 


In contrast to partial correlation approaches such as the PC-simple algorithm, the proposed method takes full advantage of signal strength, as SPACs incorporate the magnitude of coefficients. 
This is also reflected in the numerical studies where the SPAC-ALasso and the PC-simple algorithm both produce relatively small false negative rates but large false positive rates, since they tend to select fewer covariates compared with other methods. However, as signal strength increases, the false positive rate of SPAC-ALasso decreases significantly compared to the PC-simple algorithm.
Additionally, the proposed method can still achieve sign consistency for non-Gaussian distributed covariates such as categorical covariates, where partial correlation is unable to capture the conditional independence. In simulation settings with binary covariates, the proposed method performs much better than the PC-simple algorithm in terms of overall false rate. 




Although theoretical properties on the consistency of the SPAC-ALasso are not provided in this paper, the proof should be similar to that of the SPAC-Lasso.
Moreover, the SPAC idea is flexible and can be readily applied to other penalty-based methods and the generalized linear model framework.

\vskip 14pt
\noindent {\large\bf Supplementary Materials}

We provide additional conditions, theorems, tables, corollaries, and proofs for Lemma 1, all the theorems, propositions and corollaries in the online supplementary material.
\par
\vskip 14pt
\noindent {\large\bf Acknowledgements}

We would like to acknowledge support for this project
from the National Science Foundation (NSF grant DMS 1821198).
\par

\markboth{\hfill{\footnotesize\rm FEI XUE AND ANNIE QU} \hfill}
{\hfill {\footnotesize\rm SEMI-STANDARD PARTIAL COVARIANCE SELECTION} \hfill}

\bibhang=1.7pc
\bibsep=2pt
\fontsize{9}{14pt plus.8pt minus .6pt}\selectfont
\renewcommand\bibname{\large \bf References}
\expandafter\ifx\csname
natexlab\endcsname\relax\def\natexlab#1{#1}\fi
\expandafter\ifx\csname url\endcsname\relax
  \def\url#1{\texttt{#1}}\fi
\expandafter\ifx\csname urlprefix\endcsname\relax\def\urlprefix{URL}\fi

\bibliographystyle{chicago}      
\bibliography{my_bib}   

\vskip .65cm
\noindent
Department of Statistics,
Purdue University,
West Lafayette, IN 47907, USA  
\vskip 2pt
\noindent
E-mail: feixue@purdue.edu
\vskip 2pt

\noindent
Department of Statistics,
University of California Irvine,
Irvine, CA 92697, USA
\vskip 2pt
\noindent
E-mail: aqu2@uci.edu 

\end{document}